# EINSTEIN'S EQUATIONS, COSMOLOGY AND ASTROPHYSICS

Paul S. Wesson

Department of Physics and Astronomy, University of Waterloo, Waterloo, Ontario N2L 3G1, Canada

Herzberg Institute of Astrophysics, National Research Council, Victoria, B.C. V9E 2E7, Canada

Abstract:   I give a compact, pedagogical review of our present understanding of the universe as based on general relativity.  This includes the uniform models, with special reference to the cosmological 'constant'; and the equations for spherically-symmetric systems, in a particularly convenient form that aids their application to astrophysics.  New ideas in research are also outlined, notably involving extra dimensions.

This article will be a chapter in the upcoming *Springer Handbook of Spacetime* (editors: A. Ashtekar, V. Petkov).  It compresses the material normally found in a one-semester course at the fourth-year undergraduate or first-year graduate level.

Email:  psw.papers@yahoo.ca

# EINSTEIN'S EQUATIONS, COSMOLOGY AND ASTROPHYSICS

1. <u>Introduction</u>

General relativity is a remarkable subject, based on a few principles yet covering a vast and intricate array of consequences. The present account is intended mainly as an overview, compact yet reasonably complete. The material of Einstein's equations, cosmology and astrophysics is treated in Sections 2, 3 and 4. Our understanding of these subjects has progressed so greatly that they have come to represent what might be termed academic industries. But while many properties of the universe are well described by general relativity, there are indications that a deeper understanding may require an extended theory, the possible nature of which will be outlined in the conclusion of Section 5.

Einstein's theory has a broad literature, but those who work with it tend to gravitate to a few books (some of which are massive enough to justify the metaphor). The ones in the bibliography have different strengths, and together cover everything that is necessary for an understanding of the basics of the theory [1]. There are also certain subjects which occur in the following sections that are discussed in books and papers of a more technical sort [2-8]. Ideas at the forefront of research are of a diverse nature, and perhaps best approached through introductory accounts [9, 10], since it is not clear where they will lead.

## 2.    Einstein's Equations

This section is devoted to the genesis and properties of the field equations. The notation is standard, so $x^{0,123}$ are the coordinates of time and ordinary space. To avoid symbolic clutter, we adopt the usual ploy of imagining that we measure time, distance and mass in units which make the speed of light $c$, Newton's constant of gravity $G$ and Planck's constant of action $h$ all equal to unity.

The so-called fundamental constants are, in fact, not very significant in their scientific content and are only constants in the sense of being useful conventions. They arise because the history of physics saw it useful to separate the things it deals with into categories, which in mechanics we label mass, length and time [6, 9, 10]. We ascribe basic units for these things, denoted in the abstract by $M$, $L$, $T$ and in practice by convenient measures like the gram, centimetre and second. The latter are obviously man-made, but so are the former. The concepts of mass, length and time are instructive, and arise because of the ways in which humans perceive the world and comprehend it by the five senses. Over centuries of research, this approach has been honed, and nowadays we take it for granted that the equations of physics should be homogeneous in their physical dimensions.

Dimensional analysis – the traditional shortcut of the physicist – is really the application of an elementary form of group theory related to the Pi Theorem. It provides a way of checking the dimensional consistency of the equations of physics under the permutations of the three base quantities $M$, $L$, $T$. Dimensional analysis does not, of course, determine the dimensionless factors which may enter a problem, such as $\pi$ or $e$. In this regard, it should be noted that the constants of physics *do* serve the useful purpose of converting a physical proportionality to an equation in numbers. To illustrate, let us consider the classical Kepler problem. In it, the Earth (mass $m$) orbits the Sun (mass $M$) with an azimuthal velocity ($v$) at a certain radial distance ($r$). The relative motion of the frames of reference of the two objects results in what historically has come to be called the centrifugal force $mv^2 / r$. This is counterbalanced by the gravitational force of attraction between the objects, which following Newton we know to be proportional to the product of the masses and the inverse square of their separation. The essential physics of the Kepler orbit is described by the proportionality $mv^2 / r \sim Mm / r^2$. However, to convert this to an *equation* we have to insert on the right-hand side an appropriate constant $G$. Its purpose is to transpose the physical characteristics of the quantities on the one side of the law to those on the other side, so it perforce has the physical dimensions of $M^{-1}L^3T^{-2}$. It is the somewhat arbitrary manner in which constants like $G$ are introduced that has led several well-known workers to regard their presence in physics as accidental. By contrast, the cancelling of the $m$ on the left-hand side of the previous relation with the $m$ on the right-hand side is *not* trivial. It is a consequence of Einstein's Equivalence Principle, to which we will return below. The simplicity of the Kepler problem, and particularly of the answer $v = \sqrt{GM / r}$, is due to this Principle. Indeed, it is the fact that its laws are independent of the mass of a test object which makes gravitation a relatively simple science.

Quantum mechanics, in distinction to gravitation, is characterized by the unit of action $h$ introduced by Planck and named after him. Both branches of science make use of $c$, the speed of light in vacuum. The complete suite of constants with their physical dimensions is thus $G = M^{-1}L^3T^{-2}$, $h = ML^2T^{-1}$, $c = LT^{-1}$. While these constants are commonplace, it is important to realize that their dimensional contents do not 'overlap': each may be set to unity by an appropriate choice of units independent of the others. A corollary of this is that the mass of an object $m$ can be geometrized, if so desired, in both subjects. The appropriate lengths are $Gm/c^2$ and $h/mc$, the Schwarzschild radius and the Compton wavelength. The existence of these implies that it is possible, at least in principle, to construct a unified theory of gravitation and the interactions of particle physics which is based on geometry. It is also possible, as realized long ago by Planck and others, to use $G$, $h$ and $c$ to define a 'natural' set of units. (It is currently more common to use angular frequency than straight frequency in atomic problems, so $\hbar \equiv h/2\pi$ is the preferred unit.) The correspondence between natural or Planck units and the conventional gram, centimetre and second can be summarized as follows:

$$1 m_p \equiv \left( \frac{\hbar c}{G} \right)^{1/2} = 2.2 \times 10^{-5} g \qquad 1 g = 4.6 \times 10^4 m_p$$

$$1 l_p \equiv \left( \frac{G\hbar}{c^3} \right)^{1/2} = 1.6 \times 10^{-33} cm \qquad 1 cm = 6.3 \times 10^{32} l_p$$

$$1 t_p \equiv \left( \frac{G\hbar}{c^5} \right)^{1/2} = 5.4 \times 10^{-44} s \qquad 1 s = 1.9 \times 10^{43} t_p \quad .$$

In Plank units, all the constants $G$, $\hbar$ and $c$ become unity and they consequently disappear from the equations of physics.

In general relativity, the masses of objects are nearly always taken to be constants. It is therefore a theory of accelerations rather than forces. The Equivalence Principle, noted above, thus states that test masses accelerate in a gravitational field at the same rate, irrespective of their composition. This refers not only to chemical composition, but also to contributions to effective mass from binding energy and electromagnetic and other types of energy. For a particle, the Equivalence Principle removes the distinction which might be made between the gravitational mass (the quantity concerned in the object's gravitational field) and the inertial mass (the quantity which measures the object's energy content). For a fluid, however, it will be seen below that this distinction still exists and indeed follows from the field equations. The latter should not, of course, lead to consequences which depend on our choice of coordinates. The Principle of Covariance makes formal this arbitrariness of coordinate, and by use of tensors ensures that the theory leads to results whose context is independent of how we describe things. As in other theories, in general relativity the prime objective is often the calculation of the path of a test particle. The Geodesic Principle provides a formal scheme for doing this. The analog of the distance between two nearby points in the four dimensions of spacetime is the elemental interval $ds$, which also defines proper time. The interval can be extremized by varying it to isolate the shortest route, as in the symbolic relation $\delta\left[\int ds\right] = 0$. The result is the geodesic equation, whose four components give the equations of motion along the time and spatial axes (the time component involves the energy while the components in ordinary 3D space involve the momenta of the test particle). The three principles outlined in this paragraph, to do with Equivalence, Covariance and the Geodesic, form the basis of a theory which is both monolithic and far-reaching.

Einstein's field equations are usually presented as a match between the gravitational field and its source in matter. Some of the philosophical implications of this are still under discussion (see below), but the mathematical structure of the theory is straightforward. The interval between two nearby points in spacetime is defined via an extension of Pythagoras' theorem by $ds^2 = g_{\alpha\beta}dx^\alpha dx^\beta$, where a repeated index upstairs and downstairs is shorthand for summation over time $\left(x^0\right)$ and space $\left(x^{123}\right)$. The metric tensor $g_{\alpha\beta}$ is a 4 x 4 array of potentials, which is taken to be symmetric and so has 10 independent elements. Generally the potentials depend on space and time $g_{\alpha\beta}\left(x^\gamma\right)$, but locally they are constants whose magnitudes may be set to unity, defining flat Minkowski spacetime where the diagonal components are $\eta_{\alpha\beta}=\left(+1,-1,-1,-1\right)$. The derivatives of $g_{\alpha\beta}$ with respect to the coordinates define the useful objects named after Christoffel, $\Gamma^\alpha_{\beta\gamma}\equiv\left(g^{\alpha\delta}/2\right)\left(\partial g_{\beta\delta,\gamma}+\partial g_{\gamma\delta,\beta}-\partial g_{\beta\gamma,\delta}\right)$. Here the partial derivative is denoted by a comma, and should not be confused with the semicolon used to denote the covariant derivative, which takes into account the curvature of spacetime. (The covariant derivative of a vector, for example, is given by $V_{\alpha;\beta}=V_{\alpha,\beta}-\Gamma^\gamma_{\alpha\beta}A_\gamma$.) The Christoffel symbols figure in the geodesic equation mentioned above, which gives the acceleration of a test particle in terms of its 4-velocity $u^\alpha\equiv dx^\alpha/ds$, via $du^\gamma/ds+\Gamma^\gamma_{\alpha\beta}u^\alpha u^\beta=0$. They are also used to define the Riemann tensor $R^\alpha_{\beta\gamma\delta}$, which may be shown to encode all of the relevant in-formation about the gravitational field. However, the Riemann tensor has 20 independent components, whereas to obtain field equations to solve for the 10 elements of the metric tensor $g_{\alpha\beta}$ requires an object with the same number of components. This is provided by setting the upper index in $R^{\alpha}_{\beta\gamma\delta}$ equal to one of the lower indices, and summing, a process which produces the contracted tensor $R_{\mu\nu}$ named after Ricci. When this is again contracted by taking its product with the metric tensor in its upstairs or contravariant form, the result is $R = g^{\mu\nu} R_{\mu\nu} = R^0_0 + R^1_1 + R^2_2 + R^3_3$, the Ricci or curvature scalar. It can be thought of as a kind of measure of the average intensity of the gravitational field at a point in spacetime. Lastly, the combination $G_{\mu\nu} \equiv R_{\mu\nu} - (R/2)g_{\mu\nu}$ is of special interest because its 4D covariant divergence is zero by construction: $G^{\mu}_{\nu;\mu} = 0$. The geometrical object $G_{\mu\nu}$ is known as the Einstein tensor, and comprises the left-hand side of the field equations.

The preceding paragraph is standard material and familiar to many workers. However, it is not so widely known that Einstein wished to follow the same procedure for the *other* side of his field equations. That is, he wished to replace the common properties of matter, such as the density $\rho$ and pressure $p$, by algebraic expressions. He termed the former "base wood" and the latter "fine marble". In his later years, Einstein attempted to find a way to affect this transmutation by using an extra dimension. This had already been shown by Kaluza to unify the gravitational and electromagnetic equations of classical theory, and Kaluza suggested an extension to quantum theory that was designed to explain the magnitude of the electron charge in terms of the momentum in the fifth di-mension. Unfortunately, to make algebraic progress, Kaluza was obliged to assume that the 5D theory had functions independent of the fifth coordinate (the 'cylinder' condition), and Klein took the extra dimension to be rolled up to an unobservably small size ('compactification'). These two conditions proved to be a mathematical straightjacket for the theory, which robbed it of much of its physical power and doomed Einstein's dream of a purely geometric account of gravity and matter. It was not until 1992 that a fully general 5D theory was formulated, which explained matter as being induced in 4D by the fifth dimension. Actually, it was devised by workers trying to find a geometric rationale for rest mass, who were originally ignorant of Einstein's forgotten 'dream'. Later, however, the rediscovery of an old embedding theory of differential geometry due to Campbell showed that the 5D theory (based on the 5D Ricci tensor $R_{AB}$) contained the 4D one (based on the 4D Einstein tensor $G_{\alpha\beta}$). This approach, known as Space-Time-Matter theory, was joined in 1998 by the similar Membrane theory; and it is now acknowledged that matter can be explained in geometric terms if so desired.

General relativity, in its regular 4D form, matches the Einstein tensor $G_{\alpha\beta}$ to an object which contains the phenomenological properties of matter, the energy-momentum tensor $T_{\mu\nu}$. The form of this depends somewhat on the type of matter involved, but the latter is commonly assumed to be a perfect fluid (with an isotropic pressure and a unique density and no viscosity). Then the appropriate matter tensor may be written $T_{\mu\nu} = (\rho + p)u_\mu u_\nu - pg_{\mu\nu}$, where $u_\mu$ are the 4-velocities defined before. This form may look contrived, but it can be shown that the divergence $T^\mu_{\nu;\mu} = 0$ gives back the standard equations of motion in ordinary 3D space plus the equation of continuity (conservation of mass) for the fluid.

Before joining the parts of Einstein's equations which describe the gravitational field $(G_{\mu\nu})$ and the matter $(T_{\mu\nu})$, it is necessary to tackle the notorious problem posed by the cosmological constant $\Lambda$. The mathematical possibility of adding a term $\Lambda g_{\alpha\beta}$ to the field equations arises because the metric tensor acts like a constant under covariant differentiation $(g_{\alpha\beta;\gamma} = 0)$. The presence of such a term does not therefore upset the physical considerations used to identify the left-hand side $(G_{\mu\nu})$ and the right-hand side $(T_{\mu\nu})$ of the proposed field equations. Notwithstanding this, it *does* have physical consequences. Notably, in a 3D spherically-symmetric distribution of matter, an acceleration appears which at radius $r$ is $\Lambda r / 3$. This is a repulsion for $\Lambda > 0$, but an attraction that augments gravity if $\Lambda < 0$. Einstein strongly disliked such a $\Lambda$ term, because it acts on matter without being itself connected with matter. But Eddington, his contemporary, regarded the $\Lambda$ term as the essential foundation of cosmology, and present observations do indeed indicate its importance (see below). There has been much wrangling about the cosmological constant, both in physics and philosophy. It continues to be a subject of controversy, because certain models of elementary particles imply intense vacuum fields which correspond to a large magnitude for $\Lambda$, in apparent contradiction with astrophysical observations which imply a small, positive value for $\Lambda$ of order $10^{-56}$ cm$^{-2}$. The apparent discrepancy lies in the range $10^{80-120}$. One reasonable way of explaining this is in terms of a 5D theory, where $\Lambda$ varies with scale depending on the size of the extra di-mension (which, though, is controversial). A new angle on $\Lambda$ may actually be gained by taking from particle physics the idea that the vacuum is not merely emptiness but the seat of significant physics, and joining this to the structure necessary for a tensor-based description of gravity like general relativity. The result is that the cosmological constant may be regarded as measuring the density and pressure of the vacuum, its equation of state being $\rho_v = -p_v = +\Lambda/8\pi$. This is neat, but not without its pitfalls. For example, it is common to take the physical dimension of $\Lambda$ as $L^{-2}$, so with conventional units restored the dimensionally-correct form of the density is $\rho_v = \Lambda c^2/8\pi G$. This gives the impression that the vacuum is ultimately related to the strength of gravity, as measured by $G$. However, this is mistaken. Firstly, because there is a coupling constant $8\pi G/c^2$ in front of the energy-momentum tensor if the field equations are set up using conventional units, and this exactly cancels the similar factor in $\rho_v$ as written above. Secondly, the so-called fundamental constants are in fact disposable, as we saw before; and while it may be convenient to put them back at the end of a complicated calculation, the numerical size of a given constant depends on an arbitrary choice of units and has no real significance. By contrast, the geometrical factor in $\rho_v = \Lambda/8\pi$ does have some significance. It is composed of a conventional factor 2 connected with the standard way of expressing potentials, and a factor $4\pi$. This is connected with the fact that the surface area of a sphere of radius $r$ around a given centre in flat space is $4\pi r^2$, so the intensity of a conserved field necessarily falls off as $1/4\pi r^2$, and it is necessary to integrate over the same surface area in order to evaluate the strength of a source. This situation is identical to the one in classical electromagnetism as described by Maxwell's equations. Those equations are vector in nature, and admit of a gauge term which is the gradient of a scalar function. Similarly, while Einstein's equations are tensor in nature, they too admit of a kind of gauge term. This is just the $\Lambda g_{\mu\nu}$ discussed above. In other words, the most satisfactory way to regard the cosmological constant is in terms of a kind of gauge term for the equations of general relativity.

Putting Einstein's field equations together is now – with the knowledge of the previous discussion – a simple business. We choose to keep the $\Lambda$ term explicit, for mathematical generality and because it is indicated by modern observations. (Though it was skimped in certain older books, including a black-covered one published in 1973 that was big in size and influence.) The equations in standard form read

$$G_{\mu\nu} + \Lambda g_{\mu\nu} = 8\pi T_{\mu\nu} \qquad . \tag{1}$$

These equations, despite occupying only one line, entail a vast amount of physics. They are also remarkable in that they attempt to explain reality (as expressed by $T_{\mu\nu}$) in terms of a purely abstract quantity based on geometry (namely $G_{\mu\nu}$). While a precursor may be found in Maxwell's theory, Einstein's theory represents a fundamental break with older, mechanical ways of viewing the world. It is not the purpose of the present account to go into the many observations and tests which support the validity of the equations (1). But given the abstract mode of their genesis, it is truly remarkable that they work.

3.    Cosmology

In this section, the usual viewpoint is adopted that the universe started in some event like a big bang.  This is indicated by the traditional evidence to do with the expansion, the microwave background, and nucleosynthesis.

To these should be added the more recent evidence of the integrated radiation produced by sources like stars in galaxies [3].  These produce a background field dependent on astrophysical processes, which should not be confused with the cooled-down fireball radiation now seen as the 3K background.  The integrated 'light' from galaxies has a very low intensity, which in the optical band has only recently been constrained in a meaningful way.  It is controlled by the intensity of the sources, the redshift effect of the Hubble expansion, and the age of the universe.  The last factor is important, and models of the integrated radiation from galaxies confirm that the present age is $t_0 \simeq 13 \times 10^9$ yr approximately.  The night sky is dark because the universe has a finite past history, as expected if there was something like a big bang.

The present universe, on the basis of supernova and other data, appears to be accelerating under the influence of the cosmological constant or some similar scalar field.  There is also ample evidence from the structure of spiral galaxies, the morphology of clusters of galaxies, and the gravitational lensing of distant sources like quasars, that there is a significant density of dark matter in the universe.  The nature of this is controversial, but it could be elementary particles of some kind with a low effective temperature.  Ordinary matter, of the kind seen in stars and the optical parts of galaxies, makes up a relatively small fraction of the whole, especially in comparison to the effect of the cosmological constant regarded as a density for the vacuum (see Section 2). The relative densities of the vacuum, dark matter and ordinary matter are 74% : 22% : 4% approximately. We see that the stuff of traditional astronomy is a mere smattering.

It is difficult to match the aforementioned data to any simple model of cosmology. It is particularly difficult to find a single set of parameters which gives the evolution of the scale factor $R(t)$ as a function of cosmic time, notably in regard to the supernova data indicating acceleration at the present epoch. For this reason, the current picture is largely qualitative: following the big bang, there appears to have been a phase of rapid or inflationary expansion, with the equation of state of the vacuum $(p = -\rho)$, when the universe became relatively smooth; then there was a hot period when the equation of state of the matter was close to that of radiation $(p = \rho / 3)$; and this evolved with cooling into the later phase we observe at present, when the matter is cold and behaves like dust ($p \simeq 0$), but where the $\Lambda$-like expansion is still dominant. To model these different phases, we need to take the field equations (1) and find relevant solutions.

The required solutions are named after Friedmann, Robertson and Walker (FRW). The first reduced the field equations to a pair of convenient relations which will be examined below. The latter two workers isolated the corresponding form of the interval, which is useful for calculating distances and related quantities. The 4D interval consists of two parts: a simple time, and a measure for the 3D distance whose form ensures that all places are equivalent. The Robertson-Walker interval is given by

$$ds^2 = dt^2 - \frac{R^2(t)}{\left(1 + kr^2/4\right)^2}[dr^2 + r^2 d\Omega^2] \quad . \tag{2.1}$$

Here $d\Omega^2 \equiv \left(d\theta^2 + \sin^2\theta d\phi^2\right)$ defines the angular part of the metric in spherical polar coordinates. The radial part is expressed for ease in terms of a measure that is chosen to be comoving with the matter, which means that $r$ in (2.1) is merely a distance *label*, the same at all time for a given galaxy. The 'actual' (changing) distance involves the scale factor $R(t)$, which measures the separation between two typical galaxies at time $t$. The rate of expansion is given by Hubble's parameter $H \equiv \dot{R}/R$, where an overdot denotes the total derivative with respect to time. The second derivative of $R(t)$ is measured for historical reasons by the deceleration parameter, $q \equiv -\ddot{R}R/\dot{R}^2$. This is dimensionless, while $H$ has the units of an inverse time. (The present value of $H$ is about 70 km/s/*Mpc* in terms of its traditional but rather perverse unit, and galaxies that are not too distant recede at velocities proportional to this and the distance.) The constant $k$ in (2.1) is a normalized measure of the curvature of ordinary 3D space, and can be positive, negative or zero (see below). It should be noted that an alternative form of (2.1) appears in some texts, obtained from it by a change in the radial coordinate, thus:

$$ds^2 = c^2 dt^2 - R^2(t)\left[\frac{dr^2}{\left(1 - kr^2\right)} + r^2 d\Omega^2\right] \quad . \tag{2.2}$$

This is useful if we choose to measure $r$ from ourselves considered as 'centre', whereas (2.1) is spatially isotropic and provides a more 'global' measure. Of course, for both forms, there is no real centre and no boundary.

When the Robertson-Walker interval is used in conjunction with the Einstein field equations (1), the latter take the form of two relations which were studied by Friedmann. The assumption that the density $\rho$ and pressure $p$ of the cosmological fluid are isotropic and homogeneous (= uniform) causes the partial differential equations (1) to become ordinary differential equations in the scale factor $R(t)$ which measures the expansion. Friedmann's equations are

$$8\pi\rho = \frac{3k}{R^2} + \frac{3\dot{R}^2}{R^2} - \Lambda \tag{3.1}$$

$$8\pi p = \frac{-k}{R^2} - \frac{\dot{R}^2}{R^2} - \frac{2\ddot{R}}{R} + \Lambda \quad . \tag{3.2}$$

Here the constant $k$, as mentioned above, measures the curvature of the 3D ordinary space of the models, and is normalized to have the values $\pm 1$, $0$. It can be thought of as indicating the relative contributions of the kinetic energy and gravitational binding energy for a unit volume of the fluid. In the absence of $\Lambda$, $k = -1$ means that the balance of energies is in the direction of continued expansion, $k = +1$ means that the fluid eventually stops expanding and collapses under its own gravity, while $k = 0$ means an exact balance with a continuing but slowing expansion. However, $\Lambda$ is not absent in the real universe, which considerably complicates the dynamical solutions of (3), most of which can only be isolated by numerical means.

Some instructive things emerge from the two Friedmann equations (3) when they are combined in appropriate ways. For this, it is useful to replace $\Lambda$ by its equivalent vacuum properties (see above), and write the total density and pressure as $\rho = \rho_m + \rho_v$, $p = p_m + p_v$ with matter and vacuum parts. Then combining (3.1) with three times (3.2) to eliminate $k$ gives

$$\ddot{R} = \frac{-(4/3)\pi R^3 (\rho + 3p)}{R^2} \quad .$$

(4)

This is seen to be a quasi-Newtonian law of inverse-square attraction, when we recall that the physical distance in 3D is proportional at any time to the scale-factor $R(t)$, though this symbol does not imply a physical boundary since the cosmological fluid is continuous. It is noteworthy that the effective gravitational mass of a portion of the fluid is proportional to the combination $(\rho + 3p)$, not the Newtonian $\rho$ (which is only recovered for $p \ll \rho$). Accordingly, the combination $(\rho + 3p)$ is called the gravitational density. For pure vacuum, this combination is negative for $\Lambda > 0$ since $p = -\rho = -\Lambda / 8\pi$, and this is why a universe dominated by a positive cosmological constant experiences a cosmic repulsion. Another instructive thing emerges when the first derivative of (3.1) is combined with (3.2) to eliminate $\ddot{R}$, to give:

$$\dot{\rho} = -(\rho + p)\left(\frac{3\dot{R}}{R}\right) \quad .$$

(5)

This is seen to be a kind of stability relation for the universe, in the sense that the density adjusts in proportion to the expansion rate and the combination $(\rho + p)$. This is not a gravitational effect, and accordingly the noted combination is called the inertial density. For pure vacuum, the combination $(\rho + p)$ is zero since the equation of state is $p = -\rho$.

So the vacuum has constant density (and pressure) even though the matter in the universe is expanding.

It is apparent from the above that the universe according to Einstein can have properties quite different from those predicted by Newton. The reasons for this have primarily to do with the cosmological 'constant', the possibility that the pressure of matter may be a significant fraction of the energy density, and the fact that the speed of light is large but finite. The last of these has consequences which are subtle but ubiquitous. To briefly review these, let us temporarily reinstate conventional (non-geometrical) units for $c$. Then it is obvious that as we look to greater distances we also look back in time. Advances in observational techniques are such that we can soon expect to be able to study in detail the first generation of galaxies. At greater distances we would 'see' the primordial plasma from which the galaxies formed, and beyond that would be the zipping sea of strange particles being carried along by inflation. Since the universe is isotropic about every point and about $us$, we might in principle be able to 'see' the big-bang fireball, which would resemble a glowing shell all around us.

Horizons, however, might block our view of the remote cosmos as they do our view of the distant parts of the Earth [1]. In the cosmological context, there are actually two kinds of horizon: an event horizon separates those galaxies we can see from those we cannot ever see even as $t \rightarrow \infty$; while a particle horizon separates those galaxies we can see from those we cannot see now at $t = t_0 \simeq 13 \times 10^9$ yr. FRW models exist which have both kinds of horizon, one but not the other, or neither. To investigate these, consider the path of a photon which moves radially through a universe where distance is defined by the Robertson-Walker metric. We put $ds = 0$, $d\theta = d\phi = 0$ in (2.2), and obtain the (coordinate-based) velocity as $dr / dt = \pm c \left(1 - kr^2\right)^{1/2} / R(t)$. The sign choice here corresponds to whether the photon is moving towards or away from us. More importantly, we see that the 'speed' of the photon is *not* just $c$. It actually depends on $R(t)$, which is given by the Friedmann equations (3). This means that the distance to the particle horizon, which defines that part of the universe in causal communication with us, can be quite complicated to work out. However, algebraic expressions can be written down for the simple case where $\Lambda = 0$ and $p = 0$. Then for the three values of the curvature constant, the distances are given by:

$$d = \frac{c}{H_0 \left(2q_0 - 1\right)^{1/2}} \cos^{-1}\left(\frac{1}{q_0} - 1\right), \qquad k = +1, \qquad q_0 > \frac{1}{2}$$

$$d = \frac{2c}{H_0} = 3ct_0, \qquad\qquad\qquad k = 0, \qquad q_0 = \frac{1}{2}$$

$$d = \frac{c}{H_0 \left(1 - 2q_0\right)^{1/2}} \cosh^{-1}\left(\frac{1}{q_0} - 1\right), \qquad k = -1, \qquad q_0 < \frac{1}{2} \qquad . \qquad (6)$$

The Hubble parameter and deceleration parameter used here were defined above and are to be evaluated at the present epoch. It is apparent from these relations that the size of that part of the universe we can see is *not* just given by the product of the speed of light and the age.

The redshift $z$ is in some ways a better parameter to use as a cosmological measure than either the distance or the time. It is a parameter which is directly observable; and it runs smoothly from us ($z = 0$), through the populations of galaxies and quasars $(z \simeq 1{-}10)$, and in principle all the 'way' to the big bang $(z \to \infty)$. It is defined in terms of the scale factor of the Robertson-Walker metric at present $(t_0)$ and at emission $(t_e)$ by $1 + z \equiv R(t_0) / R(t_e)$. This neatly sidesteps long-running arguments about whether the redshift is 'caused' by the Doppler effect, gravity or some other agency, which are frame-dependent in general relativity and cannot be uniquely identified. The noted definition merely makes a statement about light waves and a ratio of scales. (It might even be imagined that the universe is momentarily static at the two instants which define the redshift, with no information available as to what happened inbetween.) Notwithstanding the utility of the redshift as a measure, it is still true that most workers have a mental picture of a universe that evolves through stages separated in time. This is actually acceptable, provided the epoch is used only in a relative sense, as an ordering device. Let us therefore return to this mode of organization, and list the solutions of the Friedmann equations (3) relevant to the successive phases of the universe.

Inflation is characterized by a rapid expansion under the influence of the cosmological constant or some similar measure of vacuum energy. The appropriate solution of (3) was found by deSitter in the early days of general relativity and is given by

$$p = -\rho = -\Lambda / 8\pi, \quad R(t) \sim e^{t/L}, \quad k = 0 \quad . \tag{7.1}$$

The length scale here is related to the cosmological constant by $\Lambda = 3/L^2$ (the proportionality sign indicates that the scale factor is arbitrary up to a constant). The present universe also appears to have a significant value of $\Lambda$, which corresponds to a length $L$ of order $10^{28}$ cm. The interval corresponding to (7.1) is

$$ds^2 = dt^2 - e^{2t/L}\left(dr^2 + r^2 d\Omega^2\right) \quad . \tag{7.2}$$

There is an alternative form of this cosmological metric, which is related by a coordinate transformation but is local in nature, thus:

$$ds^2 = \left(1 - \frac{\Lambda r^2}{3}\right)dt^2 - \frac{dr^2}{\left(1 - \Lambda r^2/3\right)} - r^2 d\Omega^2 \quad . \tag{7.3}$$

This form of the deSitter metric has been extensively used to model quantum-mechanical processes in the early universe, like tunnelling. Such processes could be of great importance if it should be shown that general relativity needs to be extended in some way. For example, it then becomes feasible to explain the big bang as a quantum event, perhaps in a higher-dimensional manifold. In this regard, it can be mentioned that (7.3) for both signs of $\Lambda$ can be embedded in a 5D manifold which is *flat*, in which case (7.3) resembles a 4D pseudosphere with radius $L$ [8]. Similarly, (7.2) can be embedded in 5D Minkowski space.

Following inflation, the universe is believed to have passed through a hot period when the matter had an equation of state similar to that of radiation. A solution of the Friedmann equation (3) has been known for a long while that has the noted properties, though it was formulated before the importance of $\Lambda$ was realized. The formal solution has

$$p = \rho / 3 = 1 / 32\pi t^2, \quad R(t) \sim t^{1/2}, \quad k = 0, \quad \Lambda = 0 \qquad . \qquad (8)$$

This solution needs to be modified as regards its global properties for $\Lambda \neq 0$, but its local properties are still those necessary for nucleosynthesis of the kind needed to explain the observed abundances of the elements.

Later, when the matter had cooled, the universe is believed to have evolved into a cold phase which persists to the present, and which is characterized by a value for the matter pressure which is effectively zero. The formal solution of (3) has

$$p = 0, \quad \rho = 1 / 6\pi t^2, \quad R(t) \sim t^{2/3}, \quad k = 0, \quad \Lambda = 0 \qquad . \qquad (9)$$

This solution, like the previous one, needs to be modified in regard to its global properties for $\Lambda \neq 0$. The solution (9) is named after Einstein/deSitter, and should not be confused with the straight deSitter solution (7.1). For many years, (9) was considered to be the closest approximation to the real universe. It is slightly ironic that modern data indicate that the old solution (7.1), with the cosmological constant so detested by Einstein, may be closer to the truth.

4.   Astrophysics

The application of general relativity to astrophysical systems is simpler than to cosmology for one main reason: the influence of the cosmological constant is negligible.

For this reason, we largely ignore it in this section. Also, despite what was stated in Section 2 about the disposability of the so-called fundamental constants, $G$ and $c$ are now made explicit in order to bring out the comparison with Newtonian theory and special relativity.

Many astrophysical systems are spherically symmetric in ordinary 3D space. The solar system is like this, though as a solution of Einstein's equations (1) it is exceptionally simple because it is approximately empty of matter except for the Sun (mass $M$). The interval may be regarded as an extended version of the local deSitter one (7.3), and is given by

$$ds^2 = \left(1 - \frac{2GM}{c^2 r} - \frac{\Lambda r^2}{3}\right)dt^2 - \frac{dr^2}{\left(1 - 2GM/c^2 r - \Lambda r^2/3\right)} - r^2 d\Omega^2 \quad . \qquad (10)$$

This is the familiar form, but it should be noted that the potential can be written $2G(M + M_v)/c^2 r$ where $M_v = (4/3)\pi r^3 \rho_v$ is the effective mass of the vacuum due to its equivalent density $\rho_v = \Lambda c^2/8\pi G$ (see Section 2). It should also be noted that while the local deSitter solution (7.3) can be embedded in flat 5D, the Schwarzschild-deSitter solution (10) cannot be embedded in a flat space of less than 6 dimensions. The fact that (10) successfully accounts for the dynamics of the solar system and binary pulsars, thereby establishing the validity of general relativity, means also that any extra dimensions must play an insignificant role in much of astrophysics.

To study other astrophysical systems where there is substantial matter, we assume the latter to be a spherically-symmetric perfect fluid described by the scalars $\rho$ and $p$ for the density and pressure. It is convenient to take the interval in the form

$$ds^2 = e^\sigma c^2 dt^2 - e^\omega dr^2 - R^2 d\Omega^2 \quad . \tag{11}$$

Here $\sigma$ and $\omega$ are metric coefficients which in general depend on the time $t$ and a radial measure $r$, which can be chosen to be comoving with the matter [1]. The latter may flow either inwards or outwards, but an element of it then maintains the same radial label $r$ (as in the Robertson-Walker metric of Section 3). By contrast, $R = R(t,r)$ is really another metric coefficient, and measures the dynamics of the fluid, though in such a way that $2\pi R$ is the circumference of a great circle around the centre of the distribution. With this setup, it is the inequality of $r$ and $R$ in (11) which characterizes the departure of ordinary 3D space from flatness due to the gravitational field of the fluid.

Given the interval (11), the question arises of how to write Einstein's equations (1) in the most informative manner. In many texts, they are written as long strings of symbols relating the derivatives of the metric coefficients $\sigma$, $\omega$, $R$ to the properties of matter $\rho$, $p$. For problems of the type being considered here, there will in general be four equations for the five unknowns. Therefore, one relation may be specified in order to balance things and hopefully find a solution (ways to do this are examined below). However, in such problems it is often useful to define a function which is first order in the derivatives as a *new* unknown, and rewrite the four *second*-order partial differential equations as five *first*-order ones [4]. For the current problem, it was found some while ago by Podurets and Misner and Sharp that the appropriate new function to define is a measure of the mass of the fluid interior to radius $r$ at time $t$, that is $m = m(r,t)$. The upshot is a set of five first-order differential equations in three metric coefficients $(\sigma, \omega, R)$ and three properties of matter $(\rho, p, m)$. Not only does this improve the tractability of the algebra, it also (after some manipulation) leads to a set of equations which have much greater physical meaning.

Writing the definition of the mass function as a relation with other quantities, the full set of field equations is:

$$\frac{2Gm}{c^2 R} = 1 + \frac{e^{-\sigma}}{c^2}\left(\frac{\partial R}{\partial t}\right)^2 - e^{-\omega}\left(\frac{\partial R}{\partial r}\right)^2 \tag{12.1}$$

$$\frac{\partial m}{\partial t} = \frac{-4\pi p R^2}{c^2}\frac{\partial R}{\partial t} \tag{12.2}$$

$$\frac{\partial m}{\partial r} = 4\pi \rho R^2 \frac{\partial R}{\partial r} \tag{12.3}$$

$$\frac{\partial \sigma}{\partial r} = \frac{-2}{p + \rho c^2}\frac{\partial p}{\partial r} \tag{12.4}$$

$$\frac{\partial \omega}{\partial t} = \frac{-2c^2}{p + \rho c^2}\frac{\partial \rho}{\partial t} - \frac{4}{R}\frac{\partial R}{\partial t} \quad . \tag{12.5}$$

The first of these equations, experience shows, is usually the hardest to solve. But it is helpful to note that it involves a balance between the Schwarzschild-like gravitational potential $Gm / c^2 R$, the kinetic energy per unit mass of the fluid $\left(\partial R / \partial t\right)^2$, and a measure of the departure of ordinary space from flatness, or equivalently the binding energy per unit mass of the fluid stored in the gravitational field $\left(\partial R / \partial r\right)^2$. The second equation above is best interpreted from right to left. It says, loosely speaking, that the force due to the pressure $p$ acting over a shell of area $4\pi R^2$ that moves at a velocity $\partial R / \partial t$ forms a quantity which in mechanics would be termed a rate of work or power, and that the mass of the fluid responds by changing at a rate consistent with Einstein's formula for the equivalent energy $mc^2$. The third equation would on integration give the usual Newtonian expression for the mass of a portion of the fluid $\left(m = 4\pi R^3 / 3\right)$ if the space were flat $\left(R = r\right)$; but since it is not, (12.3) gives the corresponding differential form for the mass of the fluid as affected by its own gravitational field. The last two equations, (12.4) and (12.5), relate the metric coefficients to the properties of the matter responsible for curving spacetime.

Solving (12) can be achieved once an extra relation is specified which balances the number of equations and the number of unknowns. There are also numerous solutions in the literature which were found by more tedious means, and whose physical meanings may be elucidated by employing (12). It would be redundant to list those solutions here, especially since reviews are available [1, 2]. The relations (12) have been applied to a wide range of problems, since they cover everything from the global cosmological fluid (the Friedmann equations are included) to tiny perturbations of it [4, 5]. Thus they lead to a more objective form of the Cosmological Principle, in which all intelligent observers judge the universe to be the same everywhere, not merely in terms of the density and pressure but in terms of *dimensionless* combinations of these and other parameters. While at the other end of the spectrum, they can be used to study the growth of material around a quantum seed to form a protogalaxy. The equations in the form (12) are especially useful in understanding the behaviour of matter under extreme circumstances, such as when the pressure approaches the energy density and the velocity of sound approaches the speed of light. New solutions like this certainly await discovery.

Ways to specify a condition which makes the set of equations (12) determinate are also various, and some examples follow:

(a) An equation of state, $p = p(\rho)$, is the traditional approach. This is particularly efficacious if information about the microscopic state of the matter is available, for example from spectral observations of a real system.

(b) Boundary conditions, in the broad sense, can help to restrict the form of a solution. These may include continuity conditions on the metric tensor if there is a join to another solution; or physical conditions, such as ones on the pressure at the centre and periphery of a system.

(c) Morphological constraints, such as self-similarity. The latter technique is especially relevant to astrophysical systems, which often lack sharp boundaries or other scales [5]. A distribution completely free of scales may be described by defining a dimensionless combined variable (say $ct/r$), so enabling the problem to be posed in ordinary rather than partial differential equations, which are easier to solve. A distribution with a single scale may be tackled using a refinement of this technique, so that problems like phase changes which involve a change in size of a physical parameter can be treated.

The preceding options are not exhaustive, and anyway there is the alternative of numerical integration. However, due to the non-linearity of Einstein's equations, an exact algebraic solution is especially valuable. The search for new solutions is left as an exercise for the motivated reader.

5.    <u>Conclusion</u>

General relativity is in the happy situation of being agreed upon by the great majority of workers and being verified by observations that stretch from the solar system to the most remote quasars.  Much of cosmology can be treated using the Friedmann equations (3) for a uniform fluid, and much of astrophysics can be handled by the more complicated equations (12) for a spherically-symmetric fluid.  In these two areas, it remains to find a single model that describes the whole history of the universe, and solutions that describe the diversity of its constituent systems.  Notwithstanding these technical shortcomings, it is still true to say that Einstein's theory provides a pretty good account of the real universe.

It is also true, however, that a shift in our understanding of the classical universe will occur if a way is found to unify it with the quantum theory of particle interactions. That a connection exists is already hinted by the cosmological-'constant' problem, wherein the energy density of the vacuum is observed to be small on macroscopic scales but inferred to be large on microscopic scales.  This problem would, of course, disappear if the properties of the vacuum prove to be variable.  But even this compromise will entail significant changes to our current accounts of both cosmology and particle physics. In fact, most workers believe that new physics will inevitably emerge from a unification of our present classical and quantum theories.  Currently, the preferred route to unification is via extra dimensions.  The basic extension is to 5D, which as mentioned before is commonly called Space-Time-Matter theory or Membrane theory, whose main concerns are with classical matter and particles, respectively.  For cosmology, perhaps the main consequence of the fifth dimension is the realization that the 4D big bang is a kind of artefact, produced by an unfortunate choice of coordinates in a flat 5D manifold. More generally, 5D relativity is a unified theory of gravitation, electromagnetism and a scalar field. It is the classical analog of the quantum interactions of the spin-2 graviton, the spin-1 photon and a spin-0 scaleron. The last may be related to the Higgs boson, which is believed to be responsible for the finite masses of other objects (though in the classical theory masses also involve shifts in spacetime along the fifth dimension). As to the old question: Why do we not 'see' the extra dimension? Well, in a way we do: it is the mass/energy all around us. This may sound strange; but adding one or more extra dimensions is actually the most effective way to extend general relativity so as to obtain new physics without upsetting established knowledge.

Our understanding of even established theory could do with some improvement, especially in what might be termed the psychology of cosmology. Anyone who has taught cosmology knows that even bright students have difficulty with the concepts raised by Einstein's theory. And even some researchers have an inadequate idea of what the big bang must have been like. This is largely because human beings are imprinted from childhood with everyday constructs which leave them ill-equipped as adults to visualize a universe without a centre or a boundary. Yet if the density and pressure depend only on the time then logic tells us that neither thing can exist. Confusion is engendered by calling the big bang an explosion, because this brings to mind a conventional bomb that sends shrapnel out from a point in 3D space until it hits some obstruction like a wall. Insofar as an analogy can be made, the big bang should be imagined as a kind of explo-sion that fills all of 3D space at the same moment, as if there is an indefinitely large number of bombs which are wired together so that they all detonate at the same instant. Even this description does not get across all of the subtleties of the Einstein singularity, but hopefully more attention will be given in the future to thinking in the 'right' way.

Acknowledgements

    Thanks go to the students who in the past asked good questions; and to the colleagues who shared their research, notably on higher-dimensional relativity (http://www.astro.uwaterloo.ca/~wesson).

References

[1]    Carrol, S.: Spacetime and Geometry: An Introduction to General Relativity. Addison-Wesley, San Francisco (2004). Rindler, W.: Relativity: Special, General and Cosmological. Oxford Un. Press, Oxford (2001). Islam, J.N.: An Introduction to Mathematical Cosmology. Cambridge Un. Press, Cambridge (1992). Misner, C.W., Thorne, K.S., Wheeler, J.A.: Gravitation. Freeman, San Francisco (1973). Robertson, H.P., Noonan, T.W.: Relativity and Cosmology. Saunders, Philadelphia (1968).

[2]    Kramer, D., Stephani, H., MacCallum, M., Herlt, E.: Exact Solutions of Einstein's Field Equations. Cambridge Un. Press, Cambridge (1980).

[3]    Overduin, J.M., Wesson, P.S.: The Light / Dark Universe. World Scientific, Singapore (2008). Overduin, J.M., Wesson, P.S.: Dark Matter and Background Light. Phys. Rep. 402, 267-406 (2004).

[4]   Wesson, P.S.: A New Look At The Cosmological Principle. Astron. Astrophys. <u>68</u>, 131-137 (1978).  Wesson, P.S.: An Exact Solution to Einstein's Equations With a Stiff Equation of State. J. Math. Phys. <u>19</u>, 2283-2284 (1978).  Wesson, P.S.: A Cosmological Solution of Einstein's Equations. J. Math. Phys. <u>25</u>, 3297-3298 (1984).  Wesson, P.S., Ponce de Leon, J.: Cosmological Solution of Einstein's Equations With Uniform Density and Nonuniform Pressure. Phys. Rev. <u>D39</u>, 420-422 (1989).  Wesson, P.S.: A Class of Solutions in General Relativity of Interest for Cosmology and Astrophysics. Astrophys. J. <u>336</u>, 58-60 (1989).

[5]   Henriksen, R.N., Emslie, A.G., Wesson, P.S.: Spacetimes with Constant Vacuum Energy Density and a Conformal Killing Vector. Phys. Rev. <u>D27</u>, 1219-1227 (1983).  Wesson, P.S.: Comments on a Class of Similarity Solutions of Einstein's Equations Relevant to the Early Universe. Phys. Rev. <u>D34</u>, 3925-3926 (1986). Wesson, P.S.: Observable Relations in an Inhomogeneous Self-Similar Cosmology. Astrophys. J. <u>228</u>, 647-663 (1979).

[6]   Petkov, V. (editor): Relativity and the Dimensionality of the World. Springer, Dordrecht (2007).  Petkov, V. (editor): Minkowski Spacetime: A Hundred Years Later. Springer, Dordrecht (2010).

[7]   Carr, B.J.: Universe or Multiverse? Cambridge Un. Press, Cambridge (2007).

[8]   Wesson, P.S.: Five-Dimensional Physics. World Scientific, Singapore (2006). Wesson, P.S.: The Geometrical Unification of Gravity with Its Source. J. Gen. Rel. Grav. <u>40</u>, 1353-1365 (2008).  Wesson, P.S., Ponce de Leon, J.: Kaluza-Klein Equa-

tions, Einstein's Equations and an Effective Energy-Momentum Tensor. J. Math. Phys. <u>33</u>, 3883-3887 (1992).

[9]    Halpern, P.: The Great Beyond: Higher Dimensions, Parallel Universes, and the Extraordinary Search for a Theory of Everything. Wiley, Hoboken (2004). Halpern, P.: The Pursuit of Destiny: A History of Prediction. Perseus, Cambridge (2000).

[10]  Halpern, P., Wesson, P.S.: Brave New Universe: Illuminating the Darkest Secrets of the Cosmos. J. Henry, Washington (2006). Wesson, P.S.: Weaving the Universe: Is Modern Cosmology Discovered Or Invented? World Scientific, Singapore (2011).